\documentclass[preprint,onecolumn,showpacs,preprintnumbers,amsmath,amssymb]{revtex4}

\usepackage{mathrsfs}
\usepackage{epsfig}
\usepackage{amsmath}
\usepackage{subfigure}

\newcommand{\bs}{\boldsymbol}
\newcommand{\be}{\begin{equation}}
\newcommand{\ee}{\end{equation}}
\newcommand{\bea}{\begin{eqnarray}}
\newcommand{\eea}{\end{eqnarray}}
\newcommand{\nn}{\nonumber}

\begin{document}

\title{Strain-mediated magnetoelectric coupling in magnetostrictive/piezoelectric heterostructures and resulting high frequency effects}

\author{K.~L.~Livesey}
\email{livesey@physics.uwa.edu.au}
\altaffiliation{Current address: Commonwealth Scientific and Industrial Research Organisation, Australian Resources Research Centre, 26 Dick Perry Ave, Kensington WA 6151, Australia}

\affiliation{School of Physics M013, University of Western Australia, 35 Stirling Hwy, Crawley WA 6009, Australia}

\date{\today}

\begin{abstract}
Magnetoelectric coupling terms are derived in piezoelectric/magnetostrictive (multiferroic) thin film heterostructures using Landau-Ginzburg free energy expansions in terms of three order parameters: strain, magnetization and electric polarization. Strain is eliminated using a particular set of interface boundary conditions. Then, a general effective medium method is used to calculate the ferromagnetic resonance frequency in a BaTiO$_3$/NiFe$_2$O$_4$ superlattice. This method differs from existing methods for treating magnetoelectric heterostructures since the magnetic and electric dipolar fields are not assumed constant but vary from one film to another. The ferromagnetic resonance frequency shift is calculated as a function of applied electric field and is compared to some experimental results. 
\end{abstract}

\pacs{75.85.+t,77.55.Nv,76.50.+g,77.55.H-}

\maketitle

%----------------------------------------------------------------------------------------------------------------------------------%
\section{Introduction}
\label{intro}

Today there is much interest in magnetoelectric heterostructures due to the possibility of creating multifunctional devices where magnetic properties can be controlled with an electric field or vice-verse \cite{NanReview}. For example, an applied electric field has been shown to shift ferromagnetic resonance (FMR) frequencies \cite{bichurin02}. This has potential application in fast microwave signal processing devices with lower energy consumption than current devices. The most promising magnetoelectric heterostructures are those containing magnetostrictive and piezoelectric phases because an interface strain-mediated magnetoelectric (ME) coupling can occur that is orders of magnitude bigger than in single-phase multiferroic materials (for a review, see Ref.~\cite{NanReview}). 

Recently we presented an effective medium method (EMM) for calculating susceptibilities and resonant frequencies in a heterostructure comprising alternating multiferroic (ferroelectric and antiferromagnet) and ferromagnetic thin films \cite{me10electromagnon}. This EMM is based on previous work on dielectric \cite{agranovich} and magnetic \cite{raj87, almeida88} heterostructures respectively in the long wavelength electrostatic/magnetostatic limit. Only dipolar coupling was considered between the films. Since strain-mediated magnetoelectric coupling in heterostructures appears to be very important, in this paper we present a theory to calculate the strain-mediated coupling so that the EMM can be applied and FMR frequency shifts can be predicted as a function of applied electric field.
 
 Existing methods to calculate the high frequency magnetic and magnetoelectric susceptibility of thin film magnetoelectric composites have focused on using a simplified EMM and solving magnetization equations of motion with a phenomenological magnetoelectric coupling term \cite{bichurin02}. The coupling term takes the form of an energy density
\be
\mathscr{E}_{ME} = B_{ikn} E_{i} M_{k} M_{n} + b_{ijkn} E_{i} E_{j} M_{k} M_{n} ,
\label{MEstrain_term}
\ee
where $\bs{E}$ is the applied electric field that is assumed \emph{constant} throughout the heterostructure and $\bs{M}$ is the magnetization in the magnetostrictive phase. The magnetoelectric coupling constants $B_{ikn}$ and $b_{ijkn}$ can be expressed as a combination of elastic, piezoelectric and magnetostrictive constants of the individual phases and are derived by considering the strain boundary conditions at the interfaces and by considering the symmetries of the individual phases \cite{bichurin01}. Bichurin \emph{et al.} have also solved simultaneous magnetization and elastic equations of motion and found enhanced magnetoelectric effects near mechanical resonances \cite{bichurin05}.

Our EMM differs from that of Bichurin \emph{et al.} in several ways. Firstly, both magnetization and dielectric equations of motion can be written and so the magnetic, magnetoelectric and electric susceptibility may all be solved for, as shown in Ref.~\cite{me10electromagnon}. Secondly, our effective medium method takes into account the interface boundary conditions for the dipolar magnetic and electric fields and these fields are not assumed constant through the heterostructure, as done in other works \cite{bichurin02,bichurin01,bichurin09}. This means that demagnetizing and depolarizing effects are included in our method and different resonant frequencies will be calculated as compared with simplified EMMs, as discussed in Ref.~\cite{me10electromagnon}. 

Interface strain-mediated magnetoelectric terms like Eq.~\eqref{MEstrain_term} can be added to the magnetic and dielectric energy densities of the heterostructure and dipolar boundary conditions together with equations of motion can be solved to find the susceptibility tensor. However, by writing the magnetoelectric energy in terms of electric field $\bs{E}$ rather than electric polarization $\bs{P}$, electric fields can drive magnetization but magnetic fields cannot drive electric polarization. We therefore aim first in Section~\ref{elim_strain} to write a strain-mediated magnetoelectric energy density in terms of $\bs{M}$ and $\bs{P}$. Writing an energy \emph{density} is not strictly valid since magnetization and electric polarization are defined in different phases of the heterostructure. But the EMM works by averaging the properties of the films so both the magnetization and polarization may be thought of as belonging to a single effective material, therefore this formalism does not present a problem.

Previously, studies have used one of two methods to derive magnetoelectric coupling in heterostructures. Either linear constituent equations are considered that relate strains to magnetic and electric fields \cite{harshe,Nan94,bichurin01} or Landau free energy expansions are made in terms of magnetization, polarization and strain order parameters \cite{WuZurb,PertsevReor, Kukhar}. In Section~\ref{elim_strain} the latter method will be used to derive the magnetoelectric coupling. The full free energy expansions for \emph{both} films will be made, which leads to very complicated equations for the magnetoelectric coupling once strain is eliminated from the equations. This is in contrast to previous works that have considered the Landau expansion for only the piezoelectric film, with a misfit strain that comes from the attached magnetostrictive film \cite{WuZurb,Kukhar}, or that have considered the expansion for only the magnetostrictive film, with a misfit strain from the piezoelectric \cite{PertsevReor}. The calculation in Section~\ref{elim_strain} therefore represents an extension of these works.

In Section~\ref{shiftFMR} it is then shown how the derived strain-mediated magnetoelectric coupling alters the FMR frequency and also may dynamically couple magnetic and electric excitations. This is done by calculating just the $zz$-component of the high-frequency magnetic susceptibility using the EMM presented in Ref.~\cite{me10electromagnon}. The FMR frequency shifts are compared to some existing experimental results.

It is hoped that the general methods presented in this work will enable experimentalist to estimate the resonant frequencies or the strain-mediated magnetoelectric coupling strength in piezoelectric/magnetostrictive heterostructures.

%----------------------------------------------------------------------------------------------------------------------------------%
\section{Strain-mediated magnetoelectric coupling for a cubic system}
\label{elim_strain}

The calculation for the magnetoelectric coupling in the piezoelectric/magnetostrictive heterostructure will proceed as follows. Landau-Ginzburg expansions for the free energy are written in terms of the order parameters: electric polarization $P_i$, magnetization $M_i$, and strains in the respective phases, $u_{ij}^{(p)}$ and $u_{ij}^{(m)}$ (where $i,j=x,y,z$). Then strain boundary conditions are imposed between the phases, the free energy is minimized with respect to strain, and finally an effective magnetoelectric coupling is derived for the heterostructure. Note that all the order parameters are considered constant within a phase. This is a good approximation for sufficiently thin films. So-called ``entire-cell" effective medium methods \cite{bob96,me06} are better adapted to treating thicker films where this assumption breaks down.

Both the magnetostrictive (denoted with a superscript $m$) and piezoelectric ($p$) free energies are written assuming that a cubic symmetry exists. While cubic symmetry often exists in the paramagnetic or paraelectric phases of these materials, usually the transition to magnetic/electric order is accompanied by a lowering of the crystal symmetry. Therefore, this treatment may at first seem like a crude approximation below the Curie temperature. In fact, this is not the case and the lowering of the crystal symmetry falls out naturally in the subsequent calculation of the strains. For example, it can be shown that assuming the electric polarization $\bs{P}$ is along the (111) direction in a bulk piezoelectric material, then the strains minimizing the free energy correspond to an elongation of the cubic cell along (111), in other words, a rhombohedral distortion or a transformation to space group R3c. (See Ref.~\cite{Amin} for a more detailed discussion.) The situation is more complex for thin films that are mechanically coupled to other materials; very different distortions may correspond to the ground state as compared with bulk crystals (see, for example, Ref.~\cite{pertsev}).

%free energy density
\subsection{Landau free energy expansions}

We start by quoting the parts of the free energy expansions for the magnetostrictive and piezoelectric phases that do not depend on strain. The magnetic part of the free energy density for the cubic magnetostrictive phase is given up to fourth order, for simplicity, by \cite{kittel49}:
\be
F_{\textrm{FM}}( a_{i} ) = K (a_{x}^{2} a_{y}^{2} + a_{y}^{2} a_{z}^{2} + a_{z}^{2} a_{x}^{2} ) - \mu_{0}M_{0} \bs{H} \cdot \bs{a},
\label{magFE}
\ee
where $K$ is the anisotropy constant, $M_{0}$ is the saturation magnetization, $\bs{H}$ is the applied magnetic field and $\bs{a}= \bs{M}/M_{0}$ is the magnetization vector normalized to unity or the magnetization direction cosines. The magnitude of the magnetization does not vary (at constant temperature) and therefore terms such as $a_{i}^{2}$ and $a_{i}^{4}$ are ignored here. The exchange energy is ignored as the magnetization is assumed to be all aligned within a film. Any dynamic motion (considered in Section~\ref{shiftFMR}) involves the magnetization in the whole film moving coherently.

In Eq.~\eqref{magFE} the demagnetizing free energy is not included. For a thin film it is equal to $\frac{1}{2} \mu_{0} M_{0}^{2} a_{z}^{2}$ but for a heterostructure it is typically less than this value and takes a non-analytic form \cite{me10electromagnon}. Here it is noted that, for the material parameters used in Section~\ref{mat_params}, the demagnetizing energy does not alter the equilibrium magnetization direction, as discussed later. The effective medium method, that will be used in Section~\ref{shiftFMR} to find the resonant frequencies, calculates the dipolar energy in an implicit and elegant way, without recourse to a computationally intensive method such as a dipole sums method, and obtains very accurate results \cite{me06}. Therefore the demagnetizing effects will be considered properly in Section~\ref{shiftFMR}.

Most piezoelectric materials (for example, BaTiO$_3$, PZT and BiFeO$_3$) undergo a first-order phase transition and therefore the free energy must have terms at least up to sixth order in $P_{i}$. Also, the magnitude of the electric polarization can vary strongly as a function of applied field $\bs{E}$. The electric part of the free energy density for the cubic piezoelectric phase is \cite{hlinka}:
\bea
F_{\textrm{FE}}( P_{i}^{} ) &=& \alpha_{1} (P_{x}^{2}+P_{y}^{2}+P_{z}^{2}) + \alpha_{11} (P_{x}^{4}+P_{y}^{4}+P_{z}^{4}) \nn\\
&& +\alpha_{12} (P_{x}^{2} P_{y}^{2} + P_{y}^{2} P_{z}^{2} + P_{z}^{2} P_{x}^{2} ) \nn \\
&& +\alpha_{111} (P_{x}^{6}+P_{y}^{6}+P_{z}^{6})  + \alpha_{123} P_{x}^{2} P_{y}^{2} P_{z}^{2} \nn \\
&& +\alpha_{112} \left[ P_{x}^{2} (P_{y}^{4}+P_{z}^{4}) + P_{y}^{2} (P_{z}^{4}+P_{x}^{4} ) \right. \nn \\
&& \left. + P_{z}^{2} (P_{x}^{4}+P_{y}^{4}) \right] - \bs{E} \cdot \bs{P} ,
\label{elecFE}
\eea
where the $\alpha$s are dielectric stiffness coefficients, measured at constant strain. A recent paper suggests that terms up to $P_{i}^{8}$ are required in order to accurately model phase transitions in BaTiO$_3$ \cite{8thOrder}. However, in the present work the temperature dependence of the polarization is not being studied and in fact terms of higher order than $P_{i}^{2}$ have little affect on the magnetoelectric coupling. Therefore, for simplicity we stick to the $P_{i}^{6}$ theory.

As in the magnetic case, the depolarizing energy (equal to $\frac{1}{2} \epsilon_{0}^{-1} P_{z}^{2}$ for a thin ferroelectric film but less than this value for a heterostructure) is ignored in Eq.~\eqref{elecFE}. It will be commented on later. Also, the gradient energy is ignored as the electric polarization is assumed to be aligned within each film. In reality, the polarization may be greatly changed near an interface but these more complicated effects are ignored here.

The elastic energy density for both cubic phases is given for simplicity by an expansion in strain up to second order terms \cite{Lovett}:
\bea
F_{\textrm{el}}( u_{ik}^{(\rho)} ) &=& \frac{1}{2}  C_{11}^{(\rho)} \left(  u_{xx}^{(\rho)2} + u_{yy}^{(\rho)2} + u_{zz}^{(\rho)2}  \right)  
\nonumber \\
&&+ \frac{1}{2} C_{44}^{(\rho)}  \left( u_{xy}^{(\rho)2} + u_{yz}^{(\rho)2} + u_{zx}^{(\rho)2}   \right)
\nonumber \\
&& + C_{12}^{(\rho)} \left( u_{xx}^{(\rho)}u_{yy}^{(\rho)} + u_{yy}^{(\rho)}u_{zz}^{(\rho)} + u_{zz}^{(\rho)}u_{xx}^{(\rho)}   \right) ,
\label{elasticFE}
\eea
where $\rho=m,p$ denotes the two phases and $u_{ij}^{(\rho)}$ are the strains. $C_{ij}^{(\rho)}$ are components of the elastic compliance tensor for each material.

The magnetostrictive energy density coupling the strains and magnetization in a cubic material is given by \cite{Birss,kittel49,kittel_phonon}:
\bea
F_{\textrm{FM-el}}( a_{i}, u_{kl}^{(m)} ) &= B_{1}^{} \left( (a_{x})^{2} u_{xx}^{(m)} + (a_{y})^{2} u_{yy}^{(m)} + (a_{z})^{2} u_{zz}^{(m)}  \right)  
\nonumber \\
&+  B_{2}^{}  \left( a_{x}a_{y} u_{xy}^{(m)} + a_{y} a_{z} u_{yz}^{(m)} + a_{z} a_{x} u_{zx}^{(m)} \right),
\label{magnetostrictiveFE}
\eea
where $B_{1}^{}$ and $B_{2}^{}$, following the notation of Kittel, are the magnetoelastic constants with units Jm$^{-3}$ and can be related to measured magnetostriction constants $\lambda_{100}$ and $\lambda_{111}$ \cite{kittel49}.

The energy density coupling strains and electric polarization is given similarly by \cite{CaoCross}
\bea
F_{\textrm{FE-el}}( P_{i}, u_{kl}^{(p)} ) &=& q_{11}^{}  \left( (P_{x})^{2} u_{xx}^{(p)} + (P_{y})^{2} u_{yy}^{(p)} + (P_{z})^{2} u_{zz}^{(p)}  \right)  
\nonumber \\
&&+ q_{12}^{} \left(  u_{xx}^{} (P_{y}^{2} + P_{z}^{2}) + u_{yy}^{} (P_{x}^{2}+P_{z}^{2}) + u_{zz} (P_{x}^{2} +P_{y}^{2})  \right)
\nonumber \\
&&+  q_{44}^{}   \left( P_{x} P_{y} u_{xy}^{(p)} + P_{y} P_{z} u_{yz}^{(p)} + P_{z} P_{x} u_{zx}^{(p)} \right),
\label{piezoelectricFE}
\eea
where $\bs{P}$ is the electric polarization and $q_{11}^{}$, $q_{12}$ and $q_{44}^{}$ are the coupling coefficients with units JmC$^{-2}$.

The total elastic contribution to the energy density of the effective medium  can be found by adding together the correctly weighted energy densities, Eqs.~\eqref{elasticFE}-\eqref{piezoelectricFE}:
\bea
F_{\mathrm{tot.el.}} ( a_{i}, P_{j}, u_{kl}^{(m)}, u_{pq}^{(p)}) &=& \frac{1}{d_{m}+d_{p}} \left\{ d_{m} \left( F_{\textrm{FM-el}}( a_{i}, u_{kl}^{(m)} ) + F_{\textrm{el}}( u_{ik}^{(m)} ) \right) \right.
\nonumber \\
&& \left. + d_{p} \left( F_{\textrm{FE-el}}( P_{i}, u_{kl}^{(p)} ) + F_{\textrm{el}} ( u_{ik}^{(p)} ) \right)  \right\} .
\label{FEtotal_hetero}
\eea
The thicknesses of the magnetostrictive and piezoelectric films are given by $d_{m}^{}$ and $d_{p}^{}$.

%boundary conditions
\subsection{Strain boundary conditions}

In order to find an approximate magnetoelectric coupling, we impose boundary conditions on the strains in the two respective phases and then minimize Eq.~\eqref{FEtotal_hetero} in order to eliminate strain from the expression. The problem then is which boundary conditions to use. Harsh\'e \emph{et al.} considered four different cases of strain boundary conditions when deriving magnetoelectric coupling between magnetic and electric fields, $\bs{H}$ and $\bs{E}$, in magnetostrictive/piezoelectric thin film composites \cite{harshe}. Here only one set of boundary conditions will be considered.

\begin{figure}[tb]
\begin{center}
\includegraphics[width=8cm]{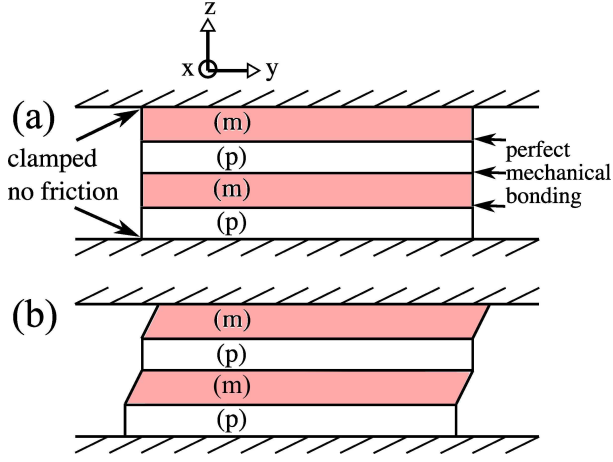}
\caption{\label{geom}(Color online) Panel (a) shows a schematic of the heterostructure geometry. The structure is clamped, with no friction, in the $z$-direction. The magnetostrictive ($m$) and piezoelectric ($p$) phases are mechanically bonded. This geometry corresponds to Case IV in Ref.~\cite{harshe}. Two repeats only are drawn for simplicity. Infinite repeats of the magnetostrictive/piezoelectric thin films are considered in the effective medium calculation in the next section. In panel (b), a schematic drawing shows how the shear strains $u_{yz}$ may be different in different films.}
\end{center}
\end{figure}

It is assumed that the heterostructure is mechanically clamped in the $z$-direction, perpendicular to the plane of each interface. This is illustrated in Fig.~\ref{geom}(a). Therefore the total distortion along the $z$ direction must be zero and the boundary condition is \cite{harshe}
\be
d_{m}^{} u_{zz}^{(m)} + d_{p}^{} u_{zz}^{(p)} =0.
\label{ZZstrainBC}
\ee
%======================%
 %======================%
 Furthermore, it is assumed that each interface has perfect mechanical bonding and therefore any distortion in the transverse directions will be equal in both phases:
\bea
u_{xx}^{(m)} &=& u_{xx}^{(p)} 
\label{XXstrainBC}\\
u_{yy}^{(m)} &=& u_{yy}^{(p)} .
\label{YYstrainBC}
\eea
This ignores any mismatch strain between the two materials due to different unit cell sizes. This can be accounted for using methods used previously for piezoelectric films \cite{pertsev} and would merely alter the boundary conditions by a constant while still ultimately giving rise to a magnetoelectric coupling.

The shear strains correspond to a change in the angle between the unstrained cartesian coordinates. Therefore, the shear in the $x-y$~plane must be equal in both films due to the perfect mechanical bonding:
\be
u_{xy}^{(m)} = u_{xy}^{(p)}.
\label{XYstrainBC}
\ee
However, the shears in the $x-z$ and $y-z$~planes have no restrictions. It is possible for a shear in the magnetostrictive material to be different than the shear in the piezoelectric material, as long as the height of the heterostructure in the $z$~direction is conserved. This is illustrated schematically in Fig.~\ref{geom}(b).

%eliminating strain
\subsection{Solutions for the strains}

Substituting Eqs.~\eqref{ZZstrainBC}-\eqref{XYstrainBC} into Eq.~\eqref{FEtotal_hetero}, $u_{xx}^{(p)}$, $u_{yy}^{(p)}$, $u_{zz}^{(p)}$ and $u_{xy}^{(p)}$ can be eliminated. Then solving $\frac{ \partial F }{ \partial u_{ij}^{(m)} } = 0$ ($i$,$j=x$,$y$,$z$) and $\frac{ \partial F }{ \partial u_{xz}^{(p)} } = 0$ and $\frac{ \partial F }{ \partial u_{yz}^{(p)} } = 0$, the strains that minimize the free energy are found. The simplest strain components to derive (since they are uncoupled from the other strains) are the shear components:
\bea
u_{xy} &=& - \frac{ d_{m} B_{2} a_{x} a_{y} + d_{p} q_{44} P_{x} P_{y} }{ d_{m}^{} C_{44}^{(m)} + d_{p}^{} C_{44}^{(p)} } 
\label{strainXY}
\\
u_{xz}^{(m)} &=& - \frac{ B_{2} a_{x} a_{z} }{ C_{44}^{(m)} }
\label{strainXZm}
\\
u_{xz}^{(p)} &=& - \frac{ q_{44} P_{x} P_{z} }{ C_{44}^{(p)} }
\label{strainXZp}
\\
u_{yz}^{(m)} &=& - \frac{ B_{2} a_{y} a_{z} }{ C_{44}^{(m)} }
\label{strainYZm}
\\
u_{yz}^{(p)} &=& - \frac{ q_{44} P_{y} P_{z} }{ C_{44}^{(p)} } .
\label{strainYZp}
\eea

By the symmetry of the boundary conditions, the transverse strains $u_{xx}$ and $u_{yy}$ are symmetric on exchange of subscripts $x$ and $y$, whereas $u_{zz}^{(m)} = - (d_{p}/d_{m}) u_{zz}^{(p)}$ is different. The transverse strains are therefore given by
\bea
u_{xx} &=& \frac{ d_{m} B_{1} f_{1}(a_{i}) + d_{p} q_{11} f_{2}(P_{j}) + d_{p} q_{12} f_{3}(P_{j})   }{ c_{1}^{} c_{2}^{2} }
\\
u_{yy} &=& u_{xx} (a_{x} \leftrightarrow a_{y}, P_{x} \leftrightarrow P_{y} )
\eea
where functions $f_1$, $f_2$ and $f_3$ are defined
\bea
f_{1}(a_{i}) &=& - a_{x}^{2} \left[ (d_{m} C_{11}^{(m)} + d_{p} C_{11}^{(p)} ) ( d_{p} C_{11}^{(m)} + d_{m} C_{11}^{(p)} ) \right. \nn \\
&& \left. + d_{m} d_{p} (C_{12}^{(m)} - C_{12}^{(p)})^2 \right] \nn \\
&&+ a_{y}^{2} \left[ (d_{m} C_{12}^{(m)} + d_{p} C_{12}^{(p)})( d_{p} C_{11}^{(m)} + d_{m} C_{11}^{(p)} ) \right. \nn \\
&& \left. - d_{m} d_{p} (C_{12}^{(m)} - C_{12}^{(p)})^2 \right] \nn \\
&&- a_{z}^{2} \phantom{}  d_{p}  (C_{12}^{(p)}-C_{12}^{(m)} ) c_{1}^{} ,
\\
f_{2}(P_{j}) &=&-  P_{x}^{2} \left[ (d_{m} C_{11}^{(m)} + d_{p} C_{11}^{(p)} ) ( d_{p} C_{11}^{(m)} + d_{m} C_{11}^{(p)} ) \right. \nn \\
&& \left. + d_{m} d_{p} (C_{12}^{(m)} - C_{12}^{(p)})^2 \right] 
\nn \\
&&+  P_{y}^{2}  \left[ (d_{m} C_{12}^{(m)} + d_{p} C_{12}^{(p)})( d_{p} C_{11}^{(m)} + d_{m} C_{11}^{(p)} ) \right. \nn \\
&& \left. - d_{m} d_{p} (C_{12}^{(m)} - C_{12}^{(p)})^2 \right] \nn \\
&&+ P_{z}^{2} \phantom{}  d_{m}  (C_{12}^{(p)}-C_{12}^{(m)} ) c_{1}^{} ,
\\
f_{3}(P_{j}) &=& P_{x}^{2} \left[ d_{m} (C_{11}^{(p)}+C_{12}^{(m)}-C_{12}^{(p)} ) ( [d_{m}-d_{p}] C_{12}^{(m)}+2 d_{p} C_{12}^{(p)} )  \right.
\nn \\
&& \left. + C_{11}^{(m)} ( d_{m} [d_{p}-d_{m}] C_{12}^{(m)} + [d_{m}^{2} + d_{p}^{2}] C_{12}^{p} ) \right]
\nn \\
&& P_{y}^{2} c_{0}^{2} - P_{z}^{2} (d_{m} C_{11}^{(p)} +d_{p} C_{11}^{(m)} ) c_{1}
\eea
and
\bea
c_{0}^{2} &=&-d_{m}^{2} C_{11}^{(m)} ( C_{11}^{(p)}+C_{12}^{(m)}-C_{12}^{(p)} ) - d_{m} d_{p} C_{11}^{(m)2}+d_{p}^{2} C_{11}^{(m)} C_{11}^{(p)} 
\nn \\
&& +d_{m} ( -d_{p} C_{11}^{(p)2} + [C_{12}^{(m)}-C_{12}^{(p)} ] [- d_{p} C_{11}^{(p)} + (d_{m}+d_{p}) C_{12}^{(m)}] \\
c_{1}^{} &=& d_{m} (C_{11}^{(m)}-C_{12}^{(m)}) + d_{p} (C_{11}^{(p)}-C_{12}^{(p)} ) , \\
c_{2}^{2}&=& \left( d_{m} C_{11}^{(p)} + d_{p} C_{11}^{(m)} \right) \nn \\
&& \times \left( d_{m} (C_{11}^{(m)}+C_{12}^{(m)}) + d_{p} (C_{11}^{(p)}+C_{12}^{(p)}) \right) \nn \\
&&- 2 d_{m} d_{p} (C_{12}^{(m)} - C_{12}^{(p)} )^2 .
\eea

The $zz$-component is given by
\bea
u_{zz}^{(m)} &=& \frac{ d_{p}^{} }{c_{2}^{2}} \left\{ -\left(m_{x}^{2}+m_{y}^{2} \right) B_{1} d_{m} \left( -C_{12}^{(m)}+C_{12}^{(p)} \right) - m_{z}^{2} B_{1} d_{p} c_{3} \right.
\nn \\
&&+ \left( P_{x}^{2} + P_{y}^{2} \right) \left( d_{p} q_{11} (C_{12}^{(m)}-C_{12}^{(p)}) + q_{12} c_{3} \right) 
\nn \\
&&\left.- P_{z}^{2}  \left( q_{11} c_{3} - 2 d_{p}^{} q_{12} (C_{12}^{(m)}-C_{12}^{(p)})  \right) \right\}
\eea
where
\bea
c_{3}^{} &=& d_{m} (C_{11}^{(m)}+C_{12}^{(m)} ) + d_{p} (C_{11}^{(p)}+C_{12}^{(p)} ) .
\label{C3function}
\eea

%Free energy with strain eliminated
\subsection{Resulting free energy}

Substituting Eqs.~\eqref{strainXY}-\eqref{C3function} for the strain back into the total strain free energy [Eq.~\eqref{FEtotal_hetero}], we obtain a energy density which takes the form
\bea
F_{\textrm{tot.el.}} (a_{i},P_{j}) &=& X_{ii}^{} a_{i}^{4} + X_{ij}^{} a_{i}^{2} a_{j}^{2} + Y_{ii}^{} P_{i}^{4}  
\nn \\
&&+ Y_{ij}^{} P_{i}^{2} P_{j}^{2}  + Z_{ii}^{} a_{i}^{2} P_{i}^{2}  
\nn \\
&&  +  Z_{ij}^{} a_{i}^{2} P_{j}^{2} + Z_{xy}^{\ast} a_{x} P_{x} a_{y} P_{y} , 
\label{FEderived}
\eea
where Einstein summation over variables $i,j=x,y,z$ is assumed and $i \ne j$. The $X$ coefficients (units Jm$^{-3}$) represent the effect of strain on the magnetization, the $Y$ coefficients (units JC$^{-4}$m$^5$) represent the effect of strain on the electric polarization, and the $Z$ coefficients (units JC$^{-2}$m) give the strength of the magnetoelectric coupling. This represents many more terms than is calculated using simpler Landau expansions, where only the complete free energy of one of the two films is considered \cite{WuZurb,PertsevReor, Kukhar}.

Due to the asymmetric boundary conditions, $X_{xx}^{} = X_{yy}^{} \ne X_{zz}$, $X_{xz}=X_{yz} \ne X_{xy}$, and so on for the other coefficients. The independent magnetic coefficients are given by
\bea
X_{xx}^{} &=& - \frac{ B_{1}^{2} d_{m}^{2} c_{4}^{2} }{ 2 (d_{m} + d_{p}) c_{1}^{} c_{2}^{2} } 
\label{Xxx}
\\
X_{zz}^{} &=& - \frac{ d_{m} d_{p} B_{1}^{2} c_{3}^{} }{2 (d_{m}+d_{p}) c_{2}^{2} } 
\eea
\bea
X_{xy}^{} &=& \frac{ d_{m}^{2} }{(d_{m} + d_{p} )} \Bigg(  \frac{B_{1}^{2} c_{5}^{2} }{c_{1}^{} c_{2}^{2}}  - \frac{ \frac{1}{2} B_{2}^{2} }{ d_{m} C_{44}^{(m)} + d_{p} C_{44}^{(p)} } \Bigg) 
\\
X_{xz}^{} &=& - \frac{d_{p} d_{m}^{2} B_{1}^{2} (C_{12}^{(p)} - C_{12}^{(m)} ) }{(d_{m} + d_{p}) c_{2}^{2} }   -   \frac{ d_{m} B_{2}^{2} }{ 2 (d_{m} + d_{p}) C_{44}^{(m)} },
\eea
where
\bea
c_{4}^{2} &=& d_{m} d_{p} C_{11}^{(m)2}  + d_{m} d_{p} \left( C_{11}^{(p)2} - (C_{12}^{(m)}-C_{12}^{(p)})^2 \right) \nn \\
&& +(d_{m}^2 + d_{p}^{2} ) C_{11}^{(m)} C_{11}^{(p)} \\
c_{5}^{2} &=& -d_{m} d_{p} (C_{12}^{(m)} - C_{12}^{(p)})^2 \nn \\
&& + (d_{m} C_{11}^{(p)} + d_{p} C_{11}^{(m)} )( d_{m} C_{12}^{m} + d_{p} C_{12}^{(p)} ) .
\eea

The independent electric coefficients are given by:
\bea
Y_{xx}^{} &=&  \frac{ - q_{11}^{2} d_{p}^{2} c_{3}^{2} + 2 q_{11} q_{12} d_{p}^{2} c_{6}^{2} + q_{12}^{2} d_{p} c_{7}^{2}  }{ 2 (d_{m} + d_{p}) c_{1}^{} c_{2}^{2} } 
\label{Yxx}
\\
Y_{zz}^{} &=& - \frac{ d_{p} \left(d_{m} q_{11}^{2} c_{3}^{} - 4 d_{m} d_{p} q_{11} q_{12} (C_{12}^{(p)}-C_{12}^{(m)}) + 2 d_{p} q_{12}^{2} (d_{m} C_{11}^{(p)}+d_{p} C_{11}^{(m)})  \right)    }{2 (d_{m}+d_{p}) c_{2}^{2} } 
\label{Yzz}
\eea
\bea
Y_{xy}^{} &=& \frac{ d_{p}^{2} q_{11}^{2} c_{5}^{2} }{ (d_{m}+d_{p}) c_{1} c_{2}^{2} } - \frac{ d_{p}^{2} q_{44}^2 }{ 2 (d_{m} + d_{p}) (d_{m} C_{44}^{(m)} + d_{p} C_{44}^{(p)} ) }
\nn \\
&& + \frac{ d_{p} q_{12}^{2} c_{8}^{3} }{ (d_{m}+d_{p}) c_{1}^{} c_{2}^{2} } + \frac{ d_{p}^2 q_{11} q_{12} c_{9}^{2} }{ (d_{m}+d_{p}) c_{1} c_{2}^{2} }
\\
Y_{xz}^{} &=& \frac{d_{p} }{(d_{m} + d_{p}) c_{2}^{2} } \left\{ d_{m}d_{p} q_{11}^{2} ( C_{12}^{(p)}-C_{12}^{(m)}) - d_{p} q_{12}^{2} \left[ d_{m} (C_{11}^{(p)} + 2 C_{12}^{(m)} - 2 C_{12}^{(p)}) + d_{p} C_{11}^{(m)} \right] \right.
\nn \\
&& \left. - q_{11}^{} q_{12}^{} \left[ ( C_{11}^{(m)} + C_{12}^{(m)} ) d_{m}^{2} + ( 2 C_{11}^{(p)}+ C_{12}^{(m)} ) d_{m} d_{p} + C_{11}^{(m)} d_{p}^{2} )  \right] \right\}
\nn \\
&& - \frac{d_{p} q_{44}^{(p)2}}{ 2 C_{44}^{(p)} (d_{m}+d_{p}) } ,
\eea
where in Eq.~\eqref{Yxx} we have constants
\bea
c_{6}^{2} &=& d_{m}^{2} \left( (C_{12}^{(m)} - C_{11}^{(m)}) (C_{12}^{(m)} - C_{12}^{(p)} ) + C_{11}^{(p)} C_{12}^{(m)} \right) \nn \\
&& + d_{m}^{} d_{p}^{} \left( C_{12}^{(m)} (3 C_{12}^{(p)} - C_{11}^{(p)} )  + 2 C_{12}^{(p)} (C_{11}^{(p)} - C_{12}^{(p)})  + C_{12}^{(m)} (C_{11}^{(m)} - C_{12}^{(m)} ) \right) \nn \\
&& + d_{p}^{2} C_{12}^{(p)} C_{11}^{(m)} \\
c_{7}^{2} &=& d_{p}^{} d_{m}^{2} \left( C_{11}^{(m)} ( 2 C_{12}^{(p)} - 3 C_{11}^{(p)} ) + C_{12}^{(m)} ( 2 C_{12}^{(m)} - 2 C_{11}^{(m)} ) \right) \nn \\
&&+ d_{m}^{} (d_{m}^{2} + d_{p}^{2} ) ( C_{12}^{(m)2} - C_{11}^{(m)2} ) - d_{p}^{3} C_{11}^{(m)} C_{11}^{(p)} \nn \\
&&- 2 d_{m}^{} d_{p}^{2} C_{11}^{(p)} ( C_{11}^{(p)} + C_{12}^{(m)} - C_{12}^{(p)} ).
\eea

The magnetoelectric coefficients, which are of most interest, are given by:
\bea
Z_{xx}^{}&=& - \frac{ d_{m} d_{p} B_{1} \left(q_{11} c_{4}^{2} - q_{12} c_{6}^{2} \right)}{ (d_{m}+ d_{p} ) c_{1}^{} c_{2}^{2} } = Z_{yy}^{} 
\\
Z_{zz}^{}&=& \frac{   d_{m} d_{p} B_{1} \left(q_{11}^{} c_{3}^{} +2 q_{12}^{} d_{p} (C_{12}^{(m)}-C_{12}^{(p)} ) \right) }{ (d_{m}+d_{p} ) c_{2}^{2} }
\\
Z_{xy}^{} &=& \frac{ d_{m} d_{p} B_{1} (q_{11} c_{5}^{2} + q_{12} c_{0}^{2} ) }{ (d_{m}+ d_{p} ) c_{1}^{} c_{2}^{2} }
\\
Z_{xz}^{} &=& - \frac{ d_{m} d_{p} B_{1}  \left( q_{11} d_{m} (C_{12}^{(m)} - C_{12}^{(p)} ) + q_{12} ( d_{m} C_{11}^{(p)} + d_{p} C_{11}^{(m)} ) \right) }{ (d_{m}+ d_{p} ) c_{2}^{2} } = Z_{yz}^{}
\\
Z_{zx}^{} &=&  \frac{  d_{m} d_{p} B_{1} \left( q_{11} d_{p} (C_{12}^{(m)} - C_{12}^{(p)} ) +q_{12} c_{3}^{}  \right) }{ (d_{m}+ d_{p} ) c_{2}^{2} }  = Z_{zy}^{}
\\
Z_{xy}^{\ast} &=& - \frac{ d_{m} d_{p} q_{44} B_{2} }{ (d_{m} + d_{p} ) ( d_{m} C_{44}^{(m)} + d_{p} C_{44}^{(p)} ) } .
\label{ZxyStar}
\eea
Notice that if the thickness of either films vanishes ($d_{m/p} \to 0$), then the magnetoelectric coupling vanishes. There is also an optimum ratio $d_{m}/d_{p}$ such that the magnetoelectric coupling is maximized, as will be shown in Section~\ref{shiftFMR}.

Now that strain has been eliminated, the effective free energy due to strain [$F_{\textrm{tot.el}}$, Eq.~\eqref{FEderived}] can be added to the unstrained free energies of the piezoelectric and magnetostrictive phases [$F_{\textrm{FM}}+F_{\textrm{FE}}$, Eqs.~\eqref{magFE} and \eqref{elecFE}]. Before calculating the ferromagnetic resonant frequency of such a system in Section~\ref{shiftFMR}, we first estimate how the strain alters the magnetic and piezoelectric constants ($K$ and $\alpha$) and estimate the strength of the magnetoelectric coupling for a BaTiO$_3$/NiFe$_2$O$_4$ heterostructure.

%BaTiO3/NiFe2O4
\subsection{BaTiO$_3$/NiFe$_2$O$_4$ coupling strength}
\label{mat_params}

A heterostructure comprising equal thickness films ($d_{m}=d_{p}$) of piezoelectric BaTiO$_3$ and magnetostrictive NiFe$_2$O$_4$ is considered. The ferroelectric parameters for BaTiO$_3$ in Eq.~\eqref{elecFE} are taken from Ref.~\cite{hlinka} and are: $\alpha_{1}=-2.772 \times 10^7$~J~m~C$^{-2}$, $\alpha_{11} = -6.476 \times 10^8$~J~m$^5$~C$^{-4}$, $\alpha_{12}=3.230 \times 10^8$~J~m$^5$~C$^{-4}$, $\alpha_{111}=8.004 \times 10^9$~J~m$^9$~C$^{-6}$, $\alpha_{112}=4.470 \times 10^9$~J~m$^9$~C$^{-6}$ and $\alpha_{123}=4.910 \times 10^9$~J~m$^9$~C$^{-6}$. The piezoelectric and elastic compliance parameters are also taken from Ref.~\cite{hlinka}: $q_{11}=-14.20 \times 10^9$~J~m~C$^{-2}$, $q_{12}=0.74 \times 10^9$~J~m~C$^{-2}$, $q_{44}=-3.14 \times 10^9$~J~m~C$^{-2}$, $C_{11}^{(p)}=27.50 \times 10^{10}$~J~m$^{-3}$, $C_{12}^{(p)}=17.90 \times10^{10}$~J~m$^{-3}$ and $C_{44}^{(p)}=5.43 \times 10^{10}$~J~m$^{-3}$.

The parameters for NiFe$_2$O$_4$ are taken from Ref.~\cite{bichurin01}: $C_{11}^{(m)}=21.99 \times 10^{10}$~N~m$^{-2}$, $C_{12}^{(m)}=10.94 \times 10^{10}$~N~m$^{-2}$, $C_{44}^{(m)}=8.12 \times 10^{10}$~N~m$^{-2}$ and $\mu_{0} M_{0}=0.32$~T ($4 \pi M_{0} = 3200$~G in CGS). The demagnetizing energy density is then $8.2\times 10^4$~J~m$^{-3}$. For the anisotropy constant the value found by Weisz \cite{weisz} is used, namely $K=4 \times 10^3$~J~m$^{-3}$. An applied field $\mu_{0} \bs{H}=0.1$~T~$\hat{\bs{x}}$ is also applied to ensure that a preferred direction of magnetization exists and later the equations of motion will be well-defined.

With these material parameters, the equilibrium magnetization and electric polarization can be found by calculating the free energy minima
 for each possible direction of the order parameters \cite{pertsev}. The equilibrium corresponds to the state with the lowest energy minima. The spontaneous electric polarization and magnetization are found to both lie along the $x$~axis (see Fig.~\ref{geom}), ie.
\bea
\bs{a} &=&(1, a_{y}, a_{z} ) 
\label{linearM} \\
\bs{P} &=& (P + p_{x}, p_{y}, p_{z} ) ,
\label{linearP}
\eea
where the lower case symbols $a_{y}$, $a_{z} <<1$ and $p_{i} << P$ ($i=x,y,z$) represent small dynamic quantities that will be relevant in Section~\ref{shiftFMR}. This result is obtained whether we consider the demagnetizing and depolarizing energy to be that of a thin film or that of bulk. In reality, the demagnetizing and depolarizing energy is somewhere between these two limits for the heterostructure and therefore the result holds for this geometry also. Then, by ignoring at first the contribution to the free energy due to strain [Eq.~\eqref{FEderived}] and only considering Eq.~\eqref{elecFE}, the polarization in zero applied field is \cite{Amin}
\be
P=\frac{ - \alpha_{11}^{} \pm \sqrt{  \alpha_{11}^{2} - 3 \alpha_{1}^{} \alpha_{111}^{} } }{ 3 \alpha_{111}^{} } = 0.265 \text{ Cm}^{-2} .
\label{unstrainedP}
\ee

Now it is possible to estimate the effect that the strains have on the magnetic and electric systems. Firstly, the strains alter $\alpha_{11}$ by over 10\%: $\alpha_{11}^{x} \to ( \alpha_{11} + Y_{xx} )$ and $ \alpha_{11}^{z} \to ( \alpha_{11} + Y_{zz} )$, where $Y_{xx}=-1.66 \times 10^8$~J~C$^{-4}$m$^{5}$ and $Y_{zz}=-1.03 \times 10^8$~J~C$^{-4}$m$^{5}$, using Eqs.~\eqref{Yxx} and \eqref{Yzz}. This in turn alters the value of the polarization calculated without strains in Eq.~\eqref{unstrainedP} by 8\% ($P=0.286$~C~m$^{-2}$). 

All of the strain-mediated magnetic, electric and magnetoelectric coefficients are listed in Table~\ref{values} for the material parameters listed above. The electric  coefficients are multiplied by $P^{4}=0.286^4$ and the magnetoelectric coefficients are multiplied by $P^{2}=0.286^2$ so that every term has units of energy density and their strengths can be compared.
%-----------------------------%
\begin{table}[h]
\caption{\label{values}Values of the strain-mediated magnetic ($X_{ij}$), electric ($Y_{ij}$) and magnetoelectric ($Z_{ij}$) coefficients for a BaTiO$_3$/NiFe$_2$O$_4$ thin film heterostructure with $d_{m}=d_{p}$, derived from Eqs.~\eqref{Xxx}-\eqref{ZxyStar}. $Y_{ij}$ coefficients are multiplied by $P^4=0.286^4$~C$^4$~m$^{-8}$ and $Z_{ij}$ coefficients are multiplied by $P^2$ so that all values are in units J~m$^{-3}$.}
\begin{ruledtabular}
\begin{tabular}{cccc}
$ X_{xx} $	& $X_{zz}$	& $X_{xy}$ & $X_{xz}$  \\
-9.86		& -6.64		& -39.7 		& -86.4	    \\
\hline \hline
$ Y_{xx} P^4$	& $Y_{zz} P^4$	& $Y_{xy} P^4$ & $Y_{xz} P^4$  \\
$-1.11 \times 10^6	$	& $-7.14 \times 10^5$	& $1.27 \times 10^6 $		& -6.88 $\times 10^4$ 		   \\
\hline \hline
$ Z_{xx} P^2$	& $Z_{zz} P^2$	& $Z_{xy} P^2$ & $Z_{xz} P^2$   \\
$6.61 \times 10^3$		& $-4.35 \times 10^3$	& $-3.99 \times10^3$ 		& -524	\\
\hline \hline
 $Z_{zx} P^2$ &  $Z_{xy}^{\ast} P^2$	&& \\
 $587 $ & $4.99 \times 10^3 $ && \\
\end{tabular}
\end{ruledtabular}
\end{table}
%-----------------------------%

Strain has a large effect on the ferroelectric parameters, as illustrated by the 8\% change in $P$, but a very small effect on the magnetic parameters. This is perhaps not surprising since lattice displacements (strains) are intimately related to ferroelectricity but are not necessary for magnetic ordering. For example, the magnetic coefficients $X_{ii}$ are all on the order of 1 - 80~J~m$^{-3}$ (see Table~\ref{values}) and so strain changes the anisotropy energy by under 1\%. From now on we therefore ignore the effect of the $X_{ii}$. 

The fact that strain has a large effect on the piezoelectric phase and a relatively small effect on the magnetostrictive phase leads to a strain-mediated magnetoelectric coupling which has an intermediate strength. It is proportional to a product of magnetostrictive/piezoelectric coefficients, such as $B_{1} q_{11}$. For example, $Z_{xx} P^2 = 6.61 \times 10^3$~J~m$^{-3}$. While this value represents less than 0.3\% of the ferroelectric energy density contributions (characterized by $\alpha_{1}$), it is \emph{larger} than the magnetic anisotropy energy in an unstrained system. Therefore, we can expect in our subsequent calculation for the high frequency susceptibility that the ferromagnetic resonance (FMR) frequency will be shifted by an electric field while the affect of a magnetic field on the electric modes will be negligible.

%----------------------------------------------------------------------------------------------------------------------------------%
\section{High frequency effects of magnetoelectric coupling}
\label{shiftFMR}

To calculate the high frequency susceptibility $\hat{\chi}$ of the heterostructure, the effective medium method presented in Ref.~\cite{me10electromagnon} is used. Unlike in Ref.~\cite{me10electromagnon}, the full $6\times6$ susceptibility is not solved for. Instead, only $\chi_{zz}^{m} \equiv m_{z}^{\text{eff}} / h_{z}^{\text{eff}}$ is solved for to demonstrate the action of the magnetoelectric coupling on the FMR frequency. In Section~\ref{elim_strain} the number of repeats of the thin films is considered finite, as illustrated in Fig.~\ref{geom}. However, the effective medium method considers the number of repeats to be infinite. The results for the frequency should be accurate for a large number of unit cell repeats and will be qualitatively correct for a small number of repeats.

\subsection{Equations for the resonant frequencies}

The calculation involves solving the linearized magnetization and electric polarization equations of motion together with Maxwell's boundary conditions for the dipolar fields. The total energy density $F = F_{\textrm{FM}}+F_{\textrm{FE}}+F_{\textrm{tot.el.}}$, the sum of Eqs.~\eqref{magFE}, \eqref{elecFE} and \eqref{FEderived}, is substituted into the equations of motion (in SI units):
\bea
\frac{ d \bs{a} }{ dt} &=& \gamma \bs{a} \times \left( - \frac{1}{M} \frac{\delta F}{\delta \bs{a} } \right)
\label{Meqnmotion} \\
\frac{ d^2 \bs{P} }{ d t^2 } &=& - \epsilon_{0} f \frac{ \delta F }{ \delta \bs{P} } , 
\label{Peqnmotion}
\eea
where $\gamma$ is the gyromagnetic ratio ($\gamma=2 \pi 2.89 \times 10^{10}$~rad~Hz~T$^{-1}$) and $f$ is the effective mass term ($f=(2 \pi)^2 1.5 \times 10^{26}$~Hz$^{2}$). Then the equations are linearized using Eqs.~\eqref{linearM} and \eqref{linearP} and small oscillatory solutions are assumed ($a_{y/z} \propto  e^{-i \omega t}$ and $p_{i} \propto e^{-i \omega t}$). The resulting equations of motion for the magnetization in the magnetostrictive phase are:
\bea
- \frac{i \omega}{ \gamma} a_{y} &=& = a_{z} \left[ B_{K} + B_{0} - B_{xx}(P) - B_{xz}(P) \right] \nn \\
&&- \mu_{0} h_{z}^{(m)}  
\label{aYeqn}  \\
- \frac{i \omega}{\gamma} a_{z} &=& -a_{y} \left[ B_{k} + B_{0} - B_{xx}(P) + B_{xy}(P) \right]  \nn \\
&& - \frac{ Z_{xy}^{\ast} P}{M} p_{y} + \mu_{0} h_{y}^{(m)} ,
\label{aZeqn}
\eea
where $\bs{B}_{0}= \mu_{0} \bs{H}_{0}$ represents the static magnetic field along the $+x$ direction, $B_{K} = \frac{2 K}{M}$ is the effective anisotropy field, $B_{ij}(P) = \frac{ 2 Z_{ij} P^2 }{M}$ is the effective magnetic field due to magnetoelectric coupling and $h_{z}^{(m)}$ and $h_{x}^{(m)}$ are components of the dipolar magnetic field in the magnetostrictive material.

In Eq.~\eqref{aZeqn}, it can be seen that the magnetoelectric coupling energy  $Z_{xy}^{\ast} a_{x} a_{y} P_{x} P_{y}$ leads to a dynamic coupling between the magnetization and the $y$-component of the electric polarization in the piezoelectric phase. Moreover, the other magnetoelectric coupling terms alter the effective static magnetic field felt by the system and therefore will alter the resonant frequency.

Since $p_{y}$ is dynamically coupled to the magnetization equations of motion, we also write its equation here:
\be
- \frac{\omega^2}{\epsilon_{0} f} p_{y} = - 2 p_{y} \left[\alpha_{1} + (\alpha_{12} +Y_{xy}) P^2+ \alpha_{112} P^4 \right] - Z_{xy}^{\ast} P a_{y} + e_{y} ,
\ee
where $e_{y}$ is a component of the electric dipolar field.

Maxwell's boundary conditions between the films for the magnetic dipolar fields in the magnetostatic limit are that the in-plane component of the magnetic field and the out-of-plane component of the magnetic induction must be continuous, ie.
\bea
h_{z}^{(m)} + a_{z} M_{0} &=& h_{z}^{(p)} = \frac{C}{\mu_{0}} 
\label{BCforB} \\
h_{x/y}^{(m)} &=& h_{x/y}^{(p)} ,
\label{BCforH}
\eea
where $C$ is a constant, defined for convenience in the calculation and $\bs{h}^{(p)}$ is the magnetic dipolar field in the piezoelectric films. The susceptibility components due to the out-of-plane dipolar field $h_{z}^{(m)/(p)}$ must be calculated first to properly include the dipolar field effects due to the boundary conditions. It is this step of including Maxwell's boundary conditions \emph{before} solving for the equations of motion that sets this type of effective medium method apart from other effective medium methods and which leads to a more sophisticated model for the demagnetizing effects on the frequencies.

The $zz$ component of the magnetic susceptibility of the effective medium can be calculated by setting $h_{y}=0=e_{y}$ and by using the weighted average:
\be
\chi_{zz}^{m}\equiv \frac{m_{z}^{\text{eff}}}{h_{z}^{\text{eff}}} = \frac{d_{m} a_{z} M }{ d_{m} h_{z}^{(m)} + d_{p} h_{z}^{(p)}}.
\label{chi_formula}
\ee
Eqs.~\eqref{aYeqn}-\eqref{BCforB} can be rearranged so that $a_{z}$, $a_{y}$, $p_{y}$, $h_{z}^{(m)}$ and $h_{z}^{(p)}$ are all written in terms of the constant $C$, which therefore cancels in Eq.~\eqref{chi_formula}. The result is
\be
\chi_{zz}^{m} = \frac{d_{m} \mu_{0} M \gamma^2 \left( B_{0} + B_{K} - B_{xx}+B_{xy} + \frac{f \epsilon_{0} Z_{xy}^{\ast 2} P^2}{M (\omega^2-\omega_{T}^2) } \right)  }{ d_{m} \left( \omega^2 - \omega_{b}^{2} \right) + d_{p} ( \omega^2 - \omega_{f}^{2} ) } ,
\label{chiZZm}
\ee
where $\omega_{b}$ is the bulk ferromagnetic frequency under strain given by:
\bea
\frac{\omega_{b}^2}{\gamma^2} &=& \left(B_{0} + B_{K} - B_{xx} + B_{xy} + \frac{f \epsilon_{0} Z_{xy}^{\ast 2} P^2}{M (\omega^2-\omega_{T}^2 ) } \right) \nn \\
&& \times \left(B_{0} + B_{K} - B_{xx} - B_{xz} + \mu_{0} M \right),
\label{BulkFreq}
\eea
$\omega_{f}$ is the thin film ferromagnetic frequency under strain given by:
\bea
\frac{\omega_{f}^2}{\gamma^2} &=& \left(B_{0} + B_{K} - B_{xx} + B_{xy} + \frac{f \epsilon_{0} Z_{xy}^{\ast 2} P^2}{M (\omega^2-\omega_{T}^2 ) } \right) \nn \\
&& \times \left(B_{0} + B_{K} - B_{xx} - B_{xz}  \right),
\label{FilmFreq}
\eea
and $\omega_{T}$ is the frequency of the transverse phonon mode associated with the ferroelectric polarization, with oscillation in the $y$ direction
\be
\omega_{T}^{2} = f \epsilon_{0} 2 \left( \alpha_{1} + (\alpha_{12}+Y_{xy}) P^2 + \alpha_{112} P^4 \right).
\label{TransFreq}
\ee

It can be noticed that Eqs.~\eqref{BulkFreq} and \eqref{FilmFreq}, are implicit equations in $\omega$. However, it can be shown that $\omega_{b/f}$ (typically in the low GHz) and $\omega_{T}$ (typically in the low THz regime) are very widely spaced in frequency. Therefore, the magnetic frequencies can be approximated very well by ignoring the term proportional to $Z_{xy}^{\ast}$ in Eqs.~\eqref{BulkFreq} and \eqref{FilmFreq}.

The pole of Eq.~\eqref{chiZZm} is given by the positive solution to
\be
0=d_{m} \left( \omega^2 - \omega_{b}^{2} \right) + d_{p} ( \omega^2 - \omega_{f}^{2} )
\label{EMMfreq}
\ee 
and corresponds to the ferromagnetic resonance frequency. %new!!!!%
This equation shows how the effective medium method calculates a resonant frequency for the heterostructure which is between that of bulk $\omega_{b}$ and that of a thin film $\omega_{f}$. In Ref.~\cite{me10electromagnon} it was shown that the effective medium FMR frequency reduces to the correct results in these well-known limits.

\subsection{Electric field shift of the FMR frequency}

\begin{figure}[h]
\begin{center}
\includegraphics[width=8cm]{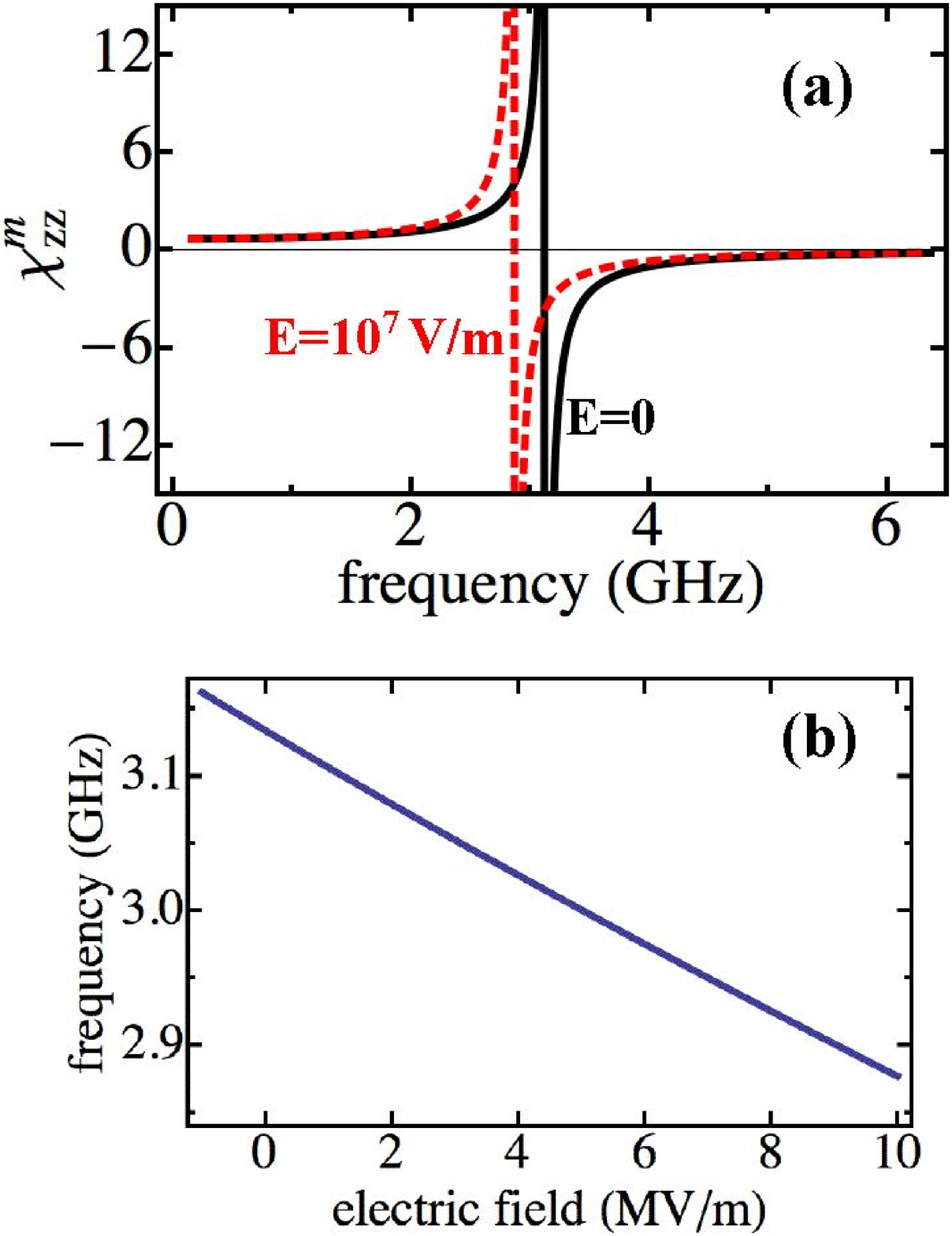}
\caption{\label{shift}(Color online) The $zz$-component of the magnetic susceptibility as a function of frequency (panel (a)) for a BaTiO$_3$/NiFe$_2$O$_4$ heterostructure, with each film of equal thickness $d_{m}=d_{p}$. A peak occurs at the ferromagnetic resonance (FMR) frequency. The peak is shifted downwards by application of an electric field in the positive $x$-direction (dashed line). In panel (b) the FMR frequency is plotted as a function of applied electric field strength.}
\end{center}
\end{figure}

Since applying an electric field alters the effective magnetic fields $B_{xx}(P)$ and $B_{xz}(P)$ in Eqs.~\eqref{BulkFreq} and \eqref{FilmFreq}, it shifts the FMR frequency of the effective medium. This is shown in Fig.~\ref{shift}. In panel (a) $\chi_{zz}^{m}$ (Eq.~\eqref{chi_formula}) is plotted as a function of frequency for no applied electric field (solid line) and for an applied electric field along the $+\hat{x}$~direction equal to $10^7$~V/m (dashed line). The pole in the susceptibility corresponds to the FMR frequency. Since damping is not included in the calculation, the peaks are not rounded at all, as is seen in real experiments. The FMR frequency is plotted as a function of electric field in panel (b). 

%======================%
 %======================%
 
 Although it is encouraging to see a shift in the FMR frequency when an electric field is applied, a large field is needed to see a 0.2~GHz shift and this appears too small for signal processing applications. However, it is slightly larger than FMR shifts seen in most recent experiments on similar systems, for example, Refs.~\cite{Tat, Das}. Alternatively, experiments may apply a constant driving frequency and sweep the applied magnetic field $H$ and then the resonant field shift as a function of $E$ may be measured. At a constant driving filed of 9~GHz, our calculation gives a shift in $H$ of 350~Oe ($\mu_{0} \Delta H=0.35$~T) when a field $E=10^8$~V/m is applied compared with when there is no field. This value is higher than measurements made, for example, by Bichurin \emph{et al.} \cite{bichurin02} on ferrite/PZT samples but on the same order of magnitude as that measued by Liu \emph{et al.} \cite{Liu}.  However, even larger effects have recently been seen when the ferrite material is replaced with the amorphous alloy FeGaB \cite{Lou08,Lou09}.

%======================%
 \begin{figure}[h]
\begin{center}
\includegraphics[width=8cm]{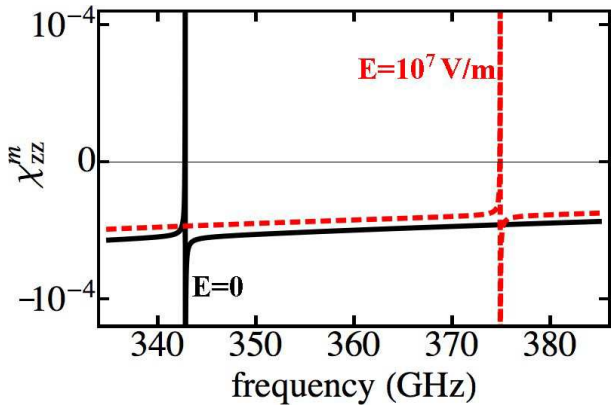}
\caption{\label{small-peak}(Color online) The $zz$-component of the magnetic susceptibility as a function of frequency for a BaTiO$_3$/NiFe$_2$O$_4$ heterostructure ($d_{m}=d_{p}$) at high frequencies. A peak occurs at the transverse phonon resonance. The peak is shifted upwards by application of a positive electric field (dashed line). Notice that the vertical scale is $10^4$ times smaller than that of Fig.~\ref{shift}(a).}
\end{center}
\end{figure}
 %======================%

It should be noted that there is a second, weak pole in the plot of $\chi_{zz}^{m}$ as a function of frequency outside the range shown in Fig.~\ref{shift}(a). This occurs at the transverse phonon mode frequency and is due to the dynamic magnetoelectric coupling discussed. It could not be predicted in an EMM where the polarization dynamics were ignored. Since it is an electric mode frequency, it is highly tunable with electric fields, as illustrated in Fig.~\ref{small-peak}. A field of $E=10^7$~V/m shifts the frequency upward by roughly 32~GHz. Although it is highly tunable, this resonance is too weak and too high to be useful for microwave applications. 
 
To try to match with particular experiments, the exact material parameters and geometry and strain boundary conditions must be input into the model. In particular, most experiments are performed with the poling direction out-of-plane whereas the results in this section have the polarization and applied electric field in-plane to simplify the effective medium method. Changing the poling direction to out-of-plane (therefore able to apply larger electric fields) and modeling specific experiments represents an exciting extension to the work presented here. 

Only one other set of strain boundary conditions is considered to see what effect different boundary conditions have on the solutions demonstrated in Figs.~\ref{shift} and \ref{small-peak}. The mechanical clamping in the $z$-direction [Eq.~\eqref{ZZstrainBC}] is relaxed and $u_{zz}^{(m)}$ and $u_{zz}^{(p)}$ are decoupled. Repeating the entire calculation for the strain-mediated magnetoelectric coupling and the resulting high frequency dynamics, it is found that the change in boundary conditions has little effect on the FMR frequency, changing the values calculated by under 0.01~GHz. However, the frequency of the transverse phonon mode, $\omega_{T}$, is shifted down in frequency by roughly 10~GHz. This reflects again the fact that the piezoelectric system is sensitive to changes in the strain whereas the magnetic system is weakly affected. Also, for this geometry, $Z_{xx}$ is the magnetoelectric coupling term which most affects the FMR frequency (see Eqs.~\eqref{BulkFreq} and \eqref{FilmFreq}) and this term is relatively unchanged when the $u_{zz}$ strains are altered. Changing the boundary conditions for in-plane strains may therefore alter the FMR frequency by a larger amount.

\subsection{Thickness effects}

\begin{figure}[h]
\begin{center}
\includegraphics[width=8cm]{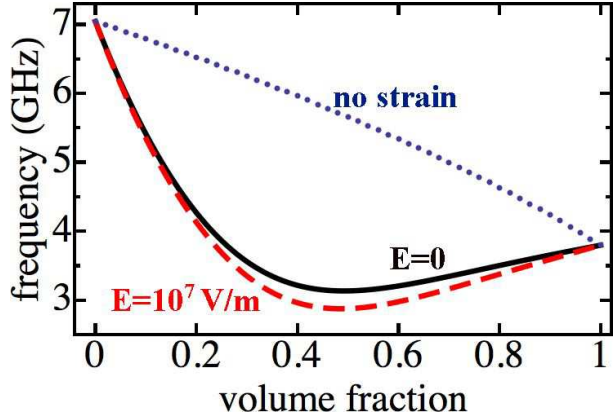}
\caption{\label{vol}(Color online) The FMR frequency, calculated using Eq.~\eqref{EMMfreq}, as a function of the magnetostrictive material's volume fraction $d_{m}/(d_{m}+d_{p})$. The solid line shows the result in zero applied electric field and the dashed line shows the result when $E=10^7$~V/m. The dotted line shows the result if there were no strain coupling to the piezoelectric.}
\end{center}
\end{figure}

Before concluding this section, the effect of the relative film thicknesses will be discussed. The ratio of the thickness of the piezoelectric versus the magnetostrictive film ($d_{p}$:$d_{m}$) changes both the magnetoelectric coupling strength \emph{and} the magnetic dipolar fields in the heterostructure. Therefore the effective medium calculation for the FMR frequency is altered through Eq.~\eqref{EMMfreq}. The result of these two effects is shown in Fig.~\ref{vol}, where the FMR frequency is plotted as a function of the volume fraction of magnetostrictive material. The solid line shows the result in zero applied electric field and the dashed line shows the result when $E=10^7$~V/m. The dotted line at the top shows the result of the effective medium calculation if the effect of strains is ignored ($B_{ij}=0$).

%======================%
 %======================%

It can be seen in Fig.~\ref{vol} that the strain coupling between the magnetostrictive and piezoelectric films changes the FMR frequency significantly. For a volume fraction of $d_{m}/(d_{m}+d_{p})=0.4$, the FMR frequency in the strained system (solid line) is half that for the unstrained system (dotted line). Fig.~\ref{vol} also reveals that a volume fraction of around 0.5 results in the largest possible change in the FMR frequency on application of an applied electric field, although this effect is weak. All three frequency calculations agree in the limits $d_{m} \to 0$ and $d_{p} \to 0$ since then the strain coupling (and electric field) have no influence on the FMR frequency. For $d_{m} \to 0$ (volume fraction is 0), the frequency of a thin magnetostrictive film is found. For $d_{p} \to 0$ (volume fraction is 1), the frequency of the bulk magnetostrictive material is found. In between, it can be seen how the effective medium method finds frequencies between these two limits since the dipolar energy is between these two extremes.

%----------------------------------------------------------------------------------------------------------------------------------%
\section{Conclusion}
\label{conclusion}

 In conclusion, it has been shown how strain-mediated magnetoelectric coupling energies can be determined for magnetostrictive/piezoelectric thin film heterostructures. This is done using Landau free energy expansions in terms of the magnetization, polarization and strains of \emph{both} thin films, and then eliminating strain using boundary conditions.
 
 The resulting terms are used to calculate the electric field shift in the FMR frequency in a BaTiO$_3$/NiFe$_2$O$_4$ heterostructure. A shift of 0.2 GHz is calculated for an applied field $E=10^7$~V/m. This is on the same order of magnitude as some experiments. The particular geometry and strain boundary conditions that are assumed may affect the strength of the derived magnetoelectric coupling terms and therefore affect the shift in the FMR frequency. An analysis of different materials will enable predictions to be made on which heterostructures are best for possible applications.
 
 What has not been mentioned so far in this work is the fact that magnetoelectric (ferrimagnetic/ferroelectric) heterostructures in which there is little strain-mediated coupling between magnetic and electric phases have also been shown to have a large magnetoelectric coupling and shifts in the FMR frequency may occur of the same magnitude as in mechanically bonded systems \cite{Ustinov,Das09}. This is due to the intrinsic hybridization of electric and magnetic fields and is not captured in our theory since the magnetostatic and electrostatic limit is assumed. Methods such as those described in Refs.~\cite{Ustinov,Demidov} may be used to calculate the FMR frequency beyond the magnetostatic limit. Future work will involve adding strain-mediated magnetoelectric couplings, like those derived in this paper, into such a theory to incorporate \emph{both} the effects of strains and hybridized fields and to determine their relative importance.

%----------------------------------------------------------------------------------------------------------------------------------% 
\begin{acknowledgments}
Support is acknowledged from the Australian Research Council. %The two referees are thanked for considerably improving the quality of the manuscript.
\end{acknowledgments}

%-------------------------------------------------%

\end{document}